\documentclass[12pt]{article}
\usepackage{epsf}
\title{ {\bf Lepton flavor violating Higgs decays induced by massive
unparticle}}
\author{\vspace{1cm}\\
        {\bf E. O. Iltan,}
        \thanks{E-mail address:
        eiltan@newton.physics.metu.edu.tr}
 \\
        Middle East Technical University, Northern Cyprus Campus,
        \\
        Guzelyurt, Mersin  10, TURKEY
\\}

\date{}

\begin{document}
\setlength{\baselineskip}{24pt}
\maketitle
\setlength{\baselineskip}{7mm}
\begin{abstract}
We predict the branching ratios of the lepton flavor violating
Higgs decays $H^0\rightarrow e^{\pm} \mu^{\pm}$, $H^0\rightarrow
e^{\pm} \tau^{\pm}$ and $H^0\rightarrow \mu^{\pm} \tau^{\pm}$ with
the assumption that lepton flavor violation is due to the
unparticle mediation. Here, we consider an effective interaction
which breaks the conformal invariance  after the electroweak
symmetry breaking and causes that unparticle becomes massive. The
new interaction results in a modification of the mediating
unparticle propagator and brings additional contribution to the
branching ratios of the lepton flavor violating decays with the
new vertex including Higgs field and two unparticle fields. We
observe that the branching ratios of the decays under
consideration lie in the range of $10^{-6}-10^{-4}$.
\end{abstract}
\thispagestyle{empty}
\newpage
\setcounter{page}{1}
%%%
%%%
%
The standard model (SM) electroweak symmetry breaking mechanism
which can explain the production of the masses of fundamental
particles will be tested at the Large Hadron Collider (LHC) and,
hopefully, the Higgs boson $H^0$, which is responsible for this
mechanism will be hunt soon. The possible decays of the Higgs
boson to the SM particles are worthwhile to study and, among them,
the lepton flavor violating (LFV) decays reach great interest
\cite{Cotti, Marfatia, Koerner, Assamagan, EoiH0unpartFV} since
the LF violation mechanism is sensitive to the physics beyond the
SM. The addition of the new Higgs doublet to the SM particle
spectrum is one of the possibility to switch on the LFV
interactions, arising from the tree level LFV couplings. In
\cite{Cotti, Marfatia, Assamagan}, $H^0\rightarrow \tau\mu$ decay
has been analyzed and the branching ratio (BR) at the order of
magnitude of $0.001-0.1$ has been estimated. In \cite{Koerner},
the observable BRs of LF changing $H^0$ decays have been obtained
in the SM with right handed neutrinos. Another possibility to
switch on the LF violation is to introduce the intermediate scalar
unparticle ($\textit{U}$) with the effective
$\textit{U}$-lepton-lepton vertex in the loop level. In \cite{
EoiH0unpartFV},  the BRs of the LFV Higgs decays $H^0\rightarrow
e^{\pm} \mu^{\pm}$, $H^0\rightarrow e^{\pm} \tau^{\pm}$ and
$H^0\rightarrow \mu^{\pm} \tau^{\pm}$ have been estimated, by
respecting the unparticle idea. Unparticles, introduced by Georgi
\cite{Georgi1, Georgi2}, come out as new degrees of freedom due to
the SM-ultraviolet sector interaction; they are massless and have
non integral scaling dimension $d_u$, around, $\Lambda_U\sim
1\,TeV$.

\begin{figure}[htb]
\vskip 3.2truein  \epsfxsize=4.5in
\leavevmode\epsffile{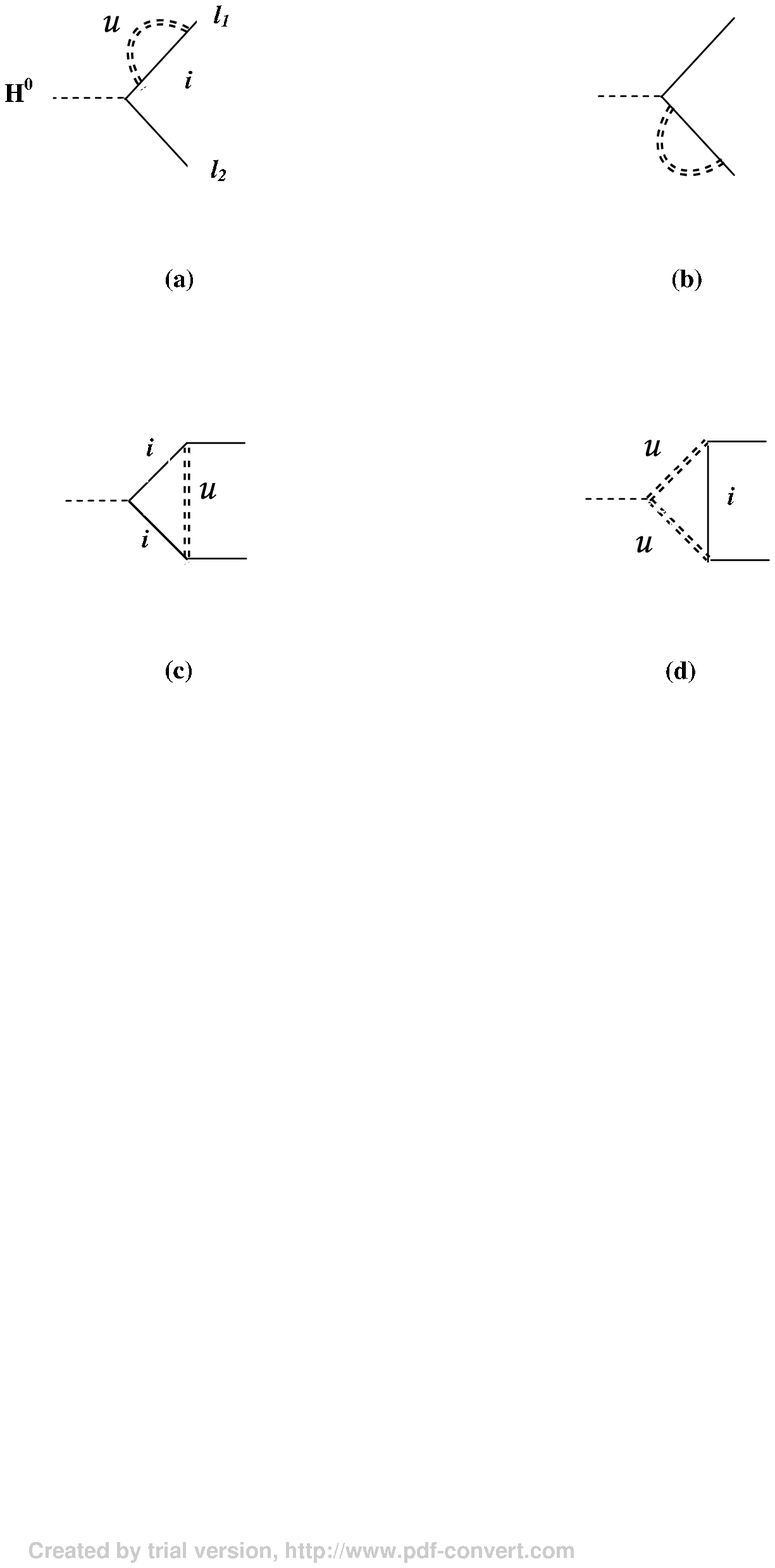} \vskip -4.0truein
\caption[]{One loop diagrams contribute to $H^0\rightarrow
l_1^-\,l_2^+$ decay with scalar unparticle mediator. Solid line
represents the lepton field: $i$ represents the internal lepton,
$l_1^-$ ($l_2^+$) outgoing lepton (anti lepton), dashed line the
Higgs field, double dashed line unparticle field.}
\label{figselfvert}
\end{figure}

In the present work we study the LFV SM Higgs decays by
considering that the LF violation exists in the one loop level and
it is carried by the effective $\textit{U}$-lepton-lepton vertex.
The effective interaction lagrangian, which is responsible for the
LFV decays, is
\begin{eqnarray}
{\cal{L}}_{FV}= \frac{1}{\Lambda_U^{du-1}}\Big (\lambda_{ij}^{S}\,
\bar{l}_{i} \,l_{j}+\lambda_{ij}^{P}\,\bar{l}_{i}
\,i\gamma_5\,l_{j}\Big)\, U \, , \label{lagrangianscalar}
\end{eqnarray}
with the lepton field $l$ and scalar (pseudoscalar) coupling
$\lambda_{ij}^{S}$ ($\lambda_{ij}^{P}$). Here we consider the
operators with the lowest possible dimension since their
contributions are dominant in the low energy effective theory (see
\cite{SChen}). Furthermore, we consider that there exists an
additional interaction which ensures a non-zero mass to unparticle
after the electroweak symmetry breaking \cite{TKikuchi} as
\begin{eqnarray}
{\cal{L}}_{U}=-
\frac{\lambda}{\Lambda_U^{2\,du-2}}\,U^2\,H^\dagger\,H
 \, ,
\label{SU}
\end{eqnarray}
and we get
\begin{eqnarray}
{\cal{L}}_{U}=-\frac{1}{2}\,
\frac{\lambda}{\Lambda_U^{2\,du-2}}\,U^2\,\Bigg(H^{0\,2}+2\,v\,H^0+v^2
\Bigg)
 \, ,
\label{SUp}
\end{eqnarray}
when the Higgs doublet develops the vacuum expectation value. The
interaction in eq.(\ref{SUp}) leads to the lagrangian
\begin{eqnarray}
{\cal{L'}}_{U}=-\frac{m_U^{4-2\,d_U}}{v}\,U^2\,H^0
 \, ,
\label{LUp}
\end{eqnarray}
with the unparticle mass
\begin{eqnarray}
m_U=\Bigg(\frac{\sqrt{\lambda}\,v}{\Lambda_U^{du-1}}
\Bigg)^{\frac{1}{2-d_U}} \, , \label{mu}
\end{eqnarray}
and this term results in an additional diagram driving the LFV
decays with the help of $\textit{U}$-lepton-lepton vertices (see
Fig.\ref{figselfvert}-(d)). Here, the non-zero unparticle mass
$m_U$ is the sign of the broken conformal invariance and one
expects that the unparticle propagator is modified. The propagator
is model dependent (see \cite{ADelgado}) and we consider the one
in the simple model \cite{PJFox, ARajaraman}
\begin{eqnarray}
\int\,d^4x\,
e^{ipx}\,<0|T\Big(U(x)\,U(0)\Big)0>=i\frac{A_{d_u}}{2\,\pi}\,
\int_0^{\infty}\,ds\,\frac{s^{d_u-2}}{p^2-\mu^2-s+i\epsilon} \, ,
\label{propagator}
\end{eqnarray}
with
\begin{eqnarray}
A_{d_u}=\frac{16\,\pi^{5/2}}{(2\,\pi)^{2\,d_u}}\,\frac{\Gamma(d_u+\frac{1}
{2})} {\Gamma(d_u-1)\,\Gamma(2\,d_u)} \, , \label{Adu}
\end{eqnarray}
and the scale $\mu$ where unparticle sector becomes a particle
sector. This choice has clues about the unparticle nature of the
hidden sector, it carries the information on the effects of the
broken scale invariance and ensures a possibility to estimate the
scale invariance breaking effects\footnote{Notice that the
modification in the propagator needs a further analysis in order
to understand whether it is based on a consistent quantum field
theory and this is beyond the scope of the present manuscript.}.
In our calculations we choose $\mu=m_U$ and $d_u\sim 1 .0$ which
is the case that unparticle behaves as if it is almost gauge
singlet scalar\footnote{ This is the case that $m_U$ lies near the
electroweak scale \cite{TKikuchi}.}.

Now, we are ready to present the BR for $H^0\rightarrow
l_1^-\,l_2^+$ decay,
\begin{eqnarray}
BR (H^0\rightarrow l_1^- \,l_2^+)=\frac{1}{16\,\pi\,m_{H^0}}\,
\frac{|M|^2}{\Gamma_{H^0}}\, , \label{BR1}
\end{eqnarray}
where $M$ is the matrix element of the LFV $H^0\rightarrow
l_1^-\,l_2^+$ decay (see Fig.\ref{figselfvert}) and $\Gamma_{H^0}$
is the Higgs total decay width. The square of the matrix element
$|M|^2$ reads
\begin{eqnarray}
|M|^2= 2\Big( m_{H^0}^2
-(m_{l_1^-}+m_{l_2^+})^2\Big)\,|A|^2+2\Big( m_{H^0}^2
-(m_{l_1^-}-m_{l_2^+})^2\Big)\,|A'|^2 \, , \label{Matrx2}
\end{eqnarray}
with the amplitudes
\begin{eqnarray}
A&=&\int^{1}_{0}\,dx\,f_{self}^S+\int^{1}_{0}\,dx\,\int^{1-x}_{0}\,dy\,
f_{vert}^S \, ,\nonumber \\
A'&=&\int^{1}_{0}\,dx\,f_{self}^{\prime\,S}+\int^{1}_{0}\,dx\,
\int^{1-x}_{0}\,dy\, f_{vert}^{\prime\,S} \, . \label{funpart}
\end{eqnarray}
The functions\footnote{$f_{self}^S$, $f_{self}^{\prime\,S}$ are
the same as the functions presented in \cite{EoiH0unpartFV} except
that the propagators $L_{self}$ and $L_{self}^{\prime}$ contain
the unparticle mass term $m_U$. On the other hand $f_{vert}^S$,
$f_{vert}^{\prime\,S}$ include additional part proportional to the
parameter $c_2$ which comes from the new interaction (see
eq.(\ref{LUp})) leading to the vertex given in
Fig.\ref{figselfvert}-d } $f_{self}^S$, $f_{self}^{\prime\,S}$,
$f_{vert}^S$, $f_{vert}^{\prime\,S}$ are
\begin{eqnarray}
f_{self}^S&=& \frac{-i\,c_1\,(1-x)^{1-d_u}}{16\,\pi^2\,\Big(
m_{l_2^+}-m_{l_1^-}\Big)\,(1-d_u)}\,\sum_{i=1}^3\,
 \Big\{(\lambda_{il_1}^S\,
\lambda_{il_2}^S+\lambda_{il_1}^P \lambda_{il_2}^P) \,
m_{l_1^-}\,m_{l_2^+}\,(1-x)\nonumber \\ &\times& \Big(
L_{self}^{d_u-1}-L_{self}^{\prime d_u-1} \Big) -
(\lambda_{il_1}^P\, \lambda_{il_2}^P-\lambda_{il_1}^S
\lambda_{il_2}^S) \,m_i\,\Big(
m_{l_2^+}\,L_{self}^{d_u-1}-m_{l_1^-}\,L_{self}^{\prime d_u-1}
\Big)
\Big\} \, , \nonumber \\ \nonumber \\
f_{self}^{\prime\,S}&=&\frac{i\,c_1\,(1-x)^{1-d_u}}{16\,\pi^2\,\Big(
m_{l_2^+}+m_{l_1^-}\Big)\,(1-d_u)}\,\sum_{i=1}^3\,
 \Big\{(\lambda_{il_1}^P\,
\lambda_{il_2}^S+\lambda_{il_1}^S \lambda_{il_2}^P) \,
m_{l_1^-}\,m_{l_2^+}\,(1-x)\nonumber \\ &\times& \Big(
L_{self}^{d_u-1}-L_{self}^{\prime d_u-1} \Big) -
(\lambda_{il_1}^P\, \lambda_{il_2}^S-\lambda_{il_1}^S
\lambda_{il_2}^P) \,m_i\,\Big(
m_{l_2^+}\,L_{self}^{d_u-1}+m_{l_1^-}\,L_{self}^{\prime d_u-1}
\Big)
\Big\} \, , \nonumber \\ \nonumber \\
f_{vert}^{S}&=& \frac{i\,c_1\,m_i\,(1-x-y)^{1-d_u}}{16\,\pi^2}\,
\sum_{i=1}^3\,\frac{1}{\,L_{vert}^{2-d_u}}\,
 \Bigg\{(\lambda_{il_1}^P\,
\lambda_{il_2}^P-\lambda_{il_1}^S \,\lambda_{il_2}^S)\,
\Big\{(1-x-y)\nonumber
\\&\times&\Bigg( m_{l_1^-}^2\,x+m_{l_2^+}^2\,y
-m_{l_2^+}\,m_{l_1^-}\Bigg)+x\,y\,m_{H^0}^2 -
\frac{2\,L_{vert}}{1-d_u}-m_i^2 \Big\} \nonumber \\&-&
(\lambda_{il_1}^P\,\lambda_{il_2}^P+\lambda_{il_1}^S
\lambda_{il_2}^S)\, m_i\,\Big(
m_{l_1^-}\,(2\,x-1)+m_{l_2^+}\,(2\,y-1)\Big) \Bigg\} \nonumber
\\&-&
\frac{i\,c_2\,\Gamma[3-2\,d_u]\,(x\,y)^{1-d_u}}{16\,\pi^2\,\Gamma[2-d_u]^2}\,
\sum_{i=1}^3\,\frac{1}{\,L_{2vert}^{3-2\,d_u}}\,
 \Bigg\{m_i\,(\lambda_{il_1}^P\,
\lambda_{il_2}^P-\lambda_{il_1}^S \,\lambda_{il_2}^S) \nonumber
\\&-&(\lambda_{il_1}^P\,\lambda_{il_2}^P+\lambda_{il_1}^S
\lambda_{il_2}^S)\,\Big( m_{l_1^-}\,x+m_{l_2^+}\,y\Big) \Bigg\} \,
, \nonumber \\
f_{vert}^{\prime\,S}&=&\frac{i\,c_1\,m_i\,(1-x-y)^{1-d_u}}{16\,\pi^2}\,
\sum_{i=1}^3\,\frac{1}{\,L_{vert}^{2-d_u}}\,
 \Bigg\{(\lambda_{il_1}^S \,\lambda_{il_2}^P-\lambda_{il_1}^P\,
\lambda_{il_2}^S)\, \Big\{(1-x-y)\nonumber
\\&\times&\Bigg( m_{l_1^-}^2\,x
+m_{l_2^+}^2\,y+m_{l_2^+}\,m_{l_1^-} \Bigg)+x\,y\,m_{H^0}^2 -
\frac{2\,L_{vert}}{1-d_u}-m_i^2 \Big\} \nonumber \\&+&
(\lambda_{il_1}^S
\lambda_{il_2}^P+\lambda_{il_1}^P\,\lambda_{il_2}^S)\, m_i\,\Big(
m_{l_1^-}\,(2\,x-1)+m_{l_2^+}\,(1-2\,y)\Big) \Bigg\}  \nonumber
\\&-&
\frac{i\,c_2\,\Gamma[3-2\,d_u]\,(x\,y)^{1-d_u}}{16\,\pi^2\,\Gamma[2-d_u]^2}\,
\sum_{i=1}^3\,\frac{1}{\,L_{2vert}^{3-2\,d_u}}\,
 \Bigg\{m_i\,(\lambda_{il_1}^S \,\lambda_{il_2}^P-\lambda_{il_1}^P\,
\lambda_{il_2}^S) \nonumber
\\&+&(\lambda_{il_1}^S
\lambda_{il_2}^P+\lambda_{il_1}^P\,\lambda_{il_2}^S)\,\Big(
m_{l_1^-}\,x-m_{l_2^+}\,y\Big) \Bigg\} \, , \label{spcouplings}
\end{eqnarray}
where $L_{self}$, $L_{self}^{\prime}$, $L_{vert}$, and $L_{2vert}$
are
\begin{eqnarray}
L_{self}&=&x\,\Big(m_{l_1^-}^2\,(1-x)-m_i^2\Big)+m_U^2\,(x-1)
\, , \nonumber \\
L_{self}^{\prime}&=&x\,\Big(m_{l_2^+}^2\,(1-x)-m_i^2\Big)+m_U^2\,(x-1)
\, ,
\nonumber \\
L_{vert}&=&(m_{l_1^-}^2\,x+m_{l_2^+}^2\,y)\,(1-x-y)-m_i^2\,(x+y)+m_{H^0}^2\,
x\,y-m_U^2\,(1-x-y) \,  ,
\nonumber \\
L_{2vert}&=&(m_{l_1^-}^2\,x+m_{l_2^+}^2\,y)\,(1-x-y)-m_i^2\,
(1-x-y)+m_{H^0}^2\,
x\,y-m_U^2\,(x+y) \, ,
\end{eqnarray}
with
\begin{eqnarray}
c_1&=&\frac{g\,A_{d_u}\,e^{-i\,\pi\,d_u}}{4\,m_W\,sin\,(d_u\pi)\,
\Lambda_u^{2\,(d_u-1)}}\,
, \nonumber \\
c_2&=&\frac{A^2_{d_u}\,m_U^{4-2\,d_u}\,e^{-2\,i\,\pi\,d_u}}{4\,v\,sin^2\,(d_u\pi)\,
\Lambda_u^{2\,(d_u-1)}}\, .
\end{eqnarray}
Here $\lambda_{il_{1(2)}}^{S,P}$ are the scalar and pseudoscalar
couplings related to the $U-i-l_1^-\,(l_2^+)$ interaction where
$i$, ($i=e,\mu,\tau$) is the internal lepton and $l_1^-\,(l_2^+)$
the outgoing lepton (anti lepton). Notice that, in the numerical
calculations,  we consider the BR due to the production of sum of
charged states, namely,
\begin{eqnarray}
BR (H^0\rightarrow l_1^{\pm}\,l_2^{\pm})=
\frac{\Gamma(H^0\rightarrow
(\bar{l}_1\,l_2+\bar{l}_2\,l_1))}{\Gamma_{H^0}} \, .\label{BR2}
\end{eqnarray}
%
%%%
%%%
%\section{Discussion}
%
\\ \\
{\Large \textbf{Discussion}}
\\ \\
This section is devoted to the analysis of the BRs of the LFV
$H^0\rightarrow l_1^- l_2^+$ decays in the case that the LF
violation is carried by the $\textit{U}$- lepton-lepton vertex.
The LFV decays exist at least in the loop level with the help of
the internal unparticle mediation. The interaction Lagrangian
given in eq.(\ref{SU}) results in a nonzero mass for unparticle
after the electroweak symmetry breaking and the propagator of
unparticle existing in the loop should be modified. In the present
work we take the propagator as (see eq.(\ref{propagator}))
\begin{eqnarray}
P(p^2)=\frac{i\,A_{d_u}}{2\,sin\,\pi\,d_u}\frac{e^{-i\,d_u\,\pi}}
{(p^2-m_U^2)^{2-d_u}} \, , \label{prop}
\end{eqnarray}
which becomes a massive scalar propagator for $d_u=1$.

The LF violation is carried by single unparticle mediation and two
unparticles mediation in the loop (see Fig.\ref{figselfvert}). The
possible two unparticles mediation brings an additional
contribution to the LFV decays with the strength which is a
function of unparticle mass $m_U$, reaching $246\,GeV$ when
$d_u\sim 1.0$ for the coupling $\lambda\sim 1.0$. In our numerical
calculations we take the scaling parameter $d_u$ not far from
$1.0$, namely $1.0\leq d_u \leq 1.2$. On the other hand we take
the coupling $\lambda$ as $\lambda\leq 1.0$ in order to guarantee
that the calculations are perturbative in case of $d_u\sim 1.0$
and we choose the energy scale $\Lambda_u$ as $\Lambda_u \sim
1.0\,(TeV)$. The FV $\textit{U}$- lepton-lepton couplings, the
scalar $\lambda^{S}_{ij}$ and pseudo scalar $\lambda^{P}_{ij}$,
are among the free parameters which we choose
$\lambda^{S}_{ij}=\lambda^{P}_{ij}=\lambda_{ij}$. Furthermore, we
first consider that the diagonal $\lambda_{ii}=\lambda_0$ and off
diagonal $\lambda_{ij}=\kappa \lambda_0, i\neq j$ couplings are
family blind with $\kappa < 1$. Second we assume that the the
diagonal couplings $\lambda_{ii}$ carry the lepton family
hierarchy, namely,
$\lambda_{\tau\tau}>\lambda_{\mu\mu}>\lambda_{ee}$, on the other
hand, the off-diagonal couplings, $\lambda_{ij}$ are family blind,
universal and  $\lambda_{ij}=\kappa \lambda_{ee}$. In our
numerical calculations, we choose $\kappa=0.5$ and we take the
magnitude of the FV coupling(s) at most $1.0$ in order to ensure
that the calculations are the perturbative for $d_u=1.0$.

In order to estimate the BR of the LFV decays under consideration
one needs the Higgs mass and its total decay width. The
theoretical upper and lower bounds of Higgs mass read $1.0\,TeV$
and $0.1\, TeV$ \cite{KHagiwara}, respectively. This is due to the
fact that one does not meet the unitarity problem and the
instability of the Higgs potential both. Furthermore, the
electroweak measurements predict the range of the Higgs mass  as
$m_{H^0}=129^{+74}_{-49}$ \cite{PDG2008} which is not in
contradiction with the theoretical results. The total Higgs decay
width is another parameter which should be restricted and it is
estimated by using the possible decays for the chosen Higgs
mass\footnote{For the light (heavy) Higgs boson, $m_{H^0} \leq 130
\, GeV$ ($m_{H^0} \sim 180\, GeV$), the leading decay mode is $b
\bar{b}$ pair \cite{Djouadi,MSpira,Drollinger} ($H^0\rightarrow W
W \rightarrow l^+ l'^- \nu_l \nu_{l'}$
\cite{RunII,Dittmar1,Dittmar2}).}. Notice that throughout our
calculations we choose $m_{H^0}=120\,(GeV)$ and we use the input
values given in Table (\ref{input}).
\begin{table}[h]
        \begin{center}
        \begin{tabular}{|l|l|}
        \hline
        \multicolumn{1}{|c|}{Parameter} &
                \multicolumn{1}{|c|}{Value}     \\
        \hline \hline
        $m_e$           & $0.0005$   (GeV)  \\
        $m_{\mu}$                   & $0.106$ (GeV) \\
        $m_{\tau}$                  & $1.780$ (GeV) \\
        $\Gamma (H^0)|_{m_{H^0}=120\,GeV}$   & $0.0029$ (GeV) \\
        $G_F$             & $1.16637 10^{-5} (GeV^{-2})$  \\
        \hline
        \end{tabular}
        \end{center}
\caption{The values of the input parameters used in the numerical
          calculations.}
\label{input}
\end{table}

\begin{figure}[htb]
\vskip -3.0truein \centering \epsfxsize=6.8in
\leavevmode\epsffile{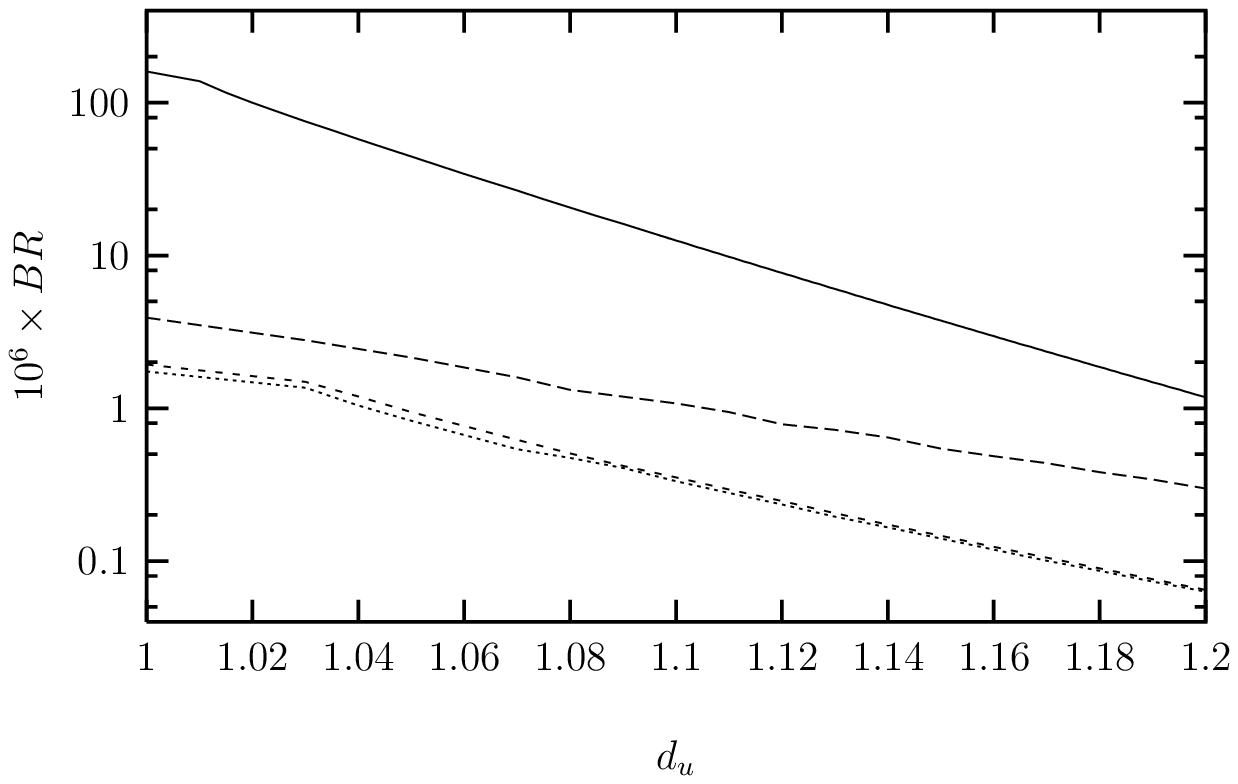} \vskip -3.0truein \caption[]{
$d_u$ dependence of the BR $(H^0\rightarrow \mu^{\pm}\, e^{\pm})$
for $\lambda_{ee}=\lambda_{\mu\mu}=\lambda_{\tau\tau}=1.0$. Here,
the solid (long dashed-short dashed-dotted) line represents the BR
for $\lambda=0.0\,(0.2-0.5-1.0)$.} \label{H0mue120du}
\end{figure}

\begin{figure}[htb]
\vskip -3.0truein \centering \epsfxsize=6.8in
\leavevmode\epsffile{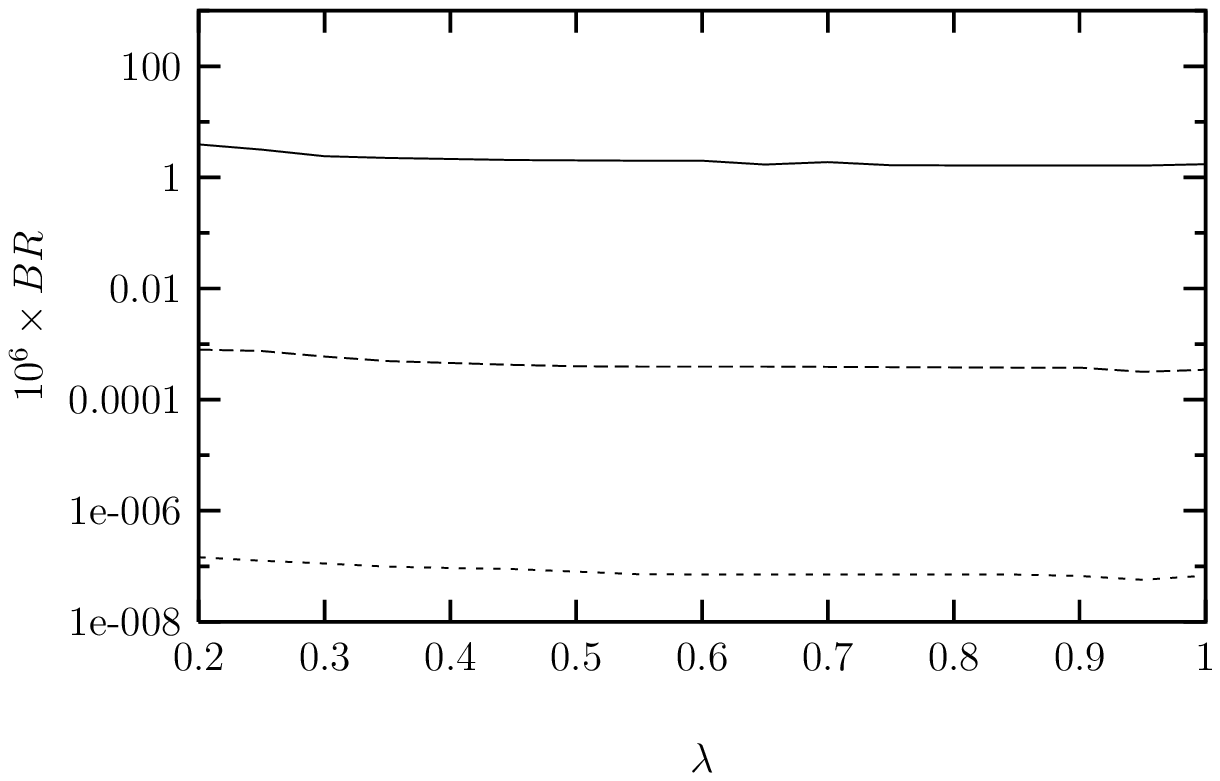} \vskip -3.0truein \caption[]{
$\lambda$ dependence of the BR $(H^0\rightarrow \mu^{\pm}\,
e^{\pm})$ for $d_u=1$. Here, the solid (long dashed-short dashed)
line represents the BR for
$\lambda_{ee}=\lambda_{\mu\mu}=\lambda_{\tau\tau}=1.0$
($\lambda_{ee}=0.1,\,
\lambda_{\mu\mu}=0.5,\,\lambda_{\tau\tau}=1.0$-$\lambda_{ee}=0.01,\,
\lambda_{\mu\mu}=0.1,\,\lambda_{\tau\tau}=1.0$).}
\label{H0mue120lam}
\end{figure}

In  Fig.\ref{H0mue120du}, we present the BR$(H^0\rightarrow
\mu^{\pm}\, e^{\pm})$ with respect to the scale parameter $d_u$
for the flavor blind (FB) couplings
$\lambda_{ee}=\lambda_{\mu\mu}=\lambda_{\tau\tau}=1.0$. Here, the
solid (long dashed-short dashed-dotted) line represents the BR for
$\lambda=0.0\,(0.2-0.5-1.0)$. The possible interaction of
unparticle with the Higgs scalar leads to a nonzero mass for
unparticle after the spontaneous symmetry breaking and the mass
term leads to a suppression in the BR. The additional term coming
from the $U-U-H^0$ vertex does not result is an enhancement in the
BR.  The BR reaches to the values of the order of $10^{-4}$ for
$\lambda=0$ and $d_u\sim 1.0$. For $\lambda \sim 1.0$ and near
$d_u\sim 1.0$ \footnote{This is the case that unparticle mass is
near the vacuum expectation value, namely $m_U\sim 246\,GeV$.} the
BR is of the order of $10^{-6}$. Fig.\ref{H0mue120lam} represents
the BR$(H^0\rightarrow \mu^{\pm}\, e^{\pm})$ with respect to
$\lambda$ for the scale parameter $d_u=1$. Here, the solid (long
dashed-short dashed) line represents the BR for
$\lambda_{ee}=\lambda_{\mu\mu}=\lambda_{\tau\tau}=1.0$
($\lambda_{ee}=0.1,\,
\lambda_{\mu\mu}=0.5,\,\lambda_{\tau\tau}=1.0$-$\lambda_{ee}=0.01,\,
\lambda_{\mu\mu}=0.1,\,\lambda_{\tau\tau}=1.0$). This figure shows
the strong sensitivity of the BR to the $U-U-H^0$ interaction
strength $\lambda$, especially for $\lambda < 0.3$.

\begin{figure}[htb]
\vskip -3.0truein \centering \epsfxsize=6.8in
\leavevmode\epsffile{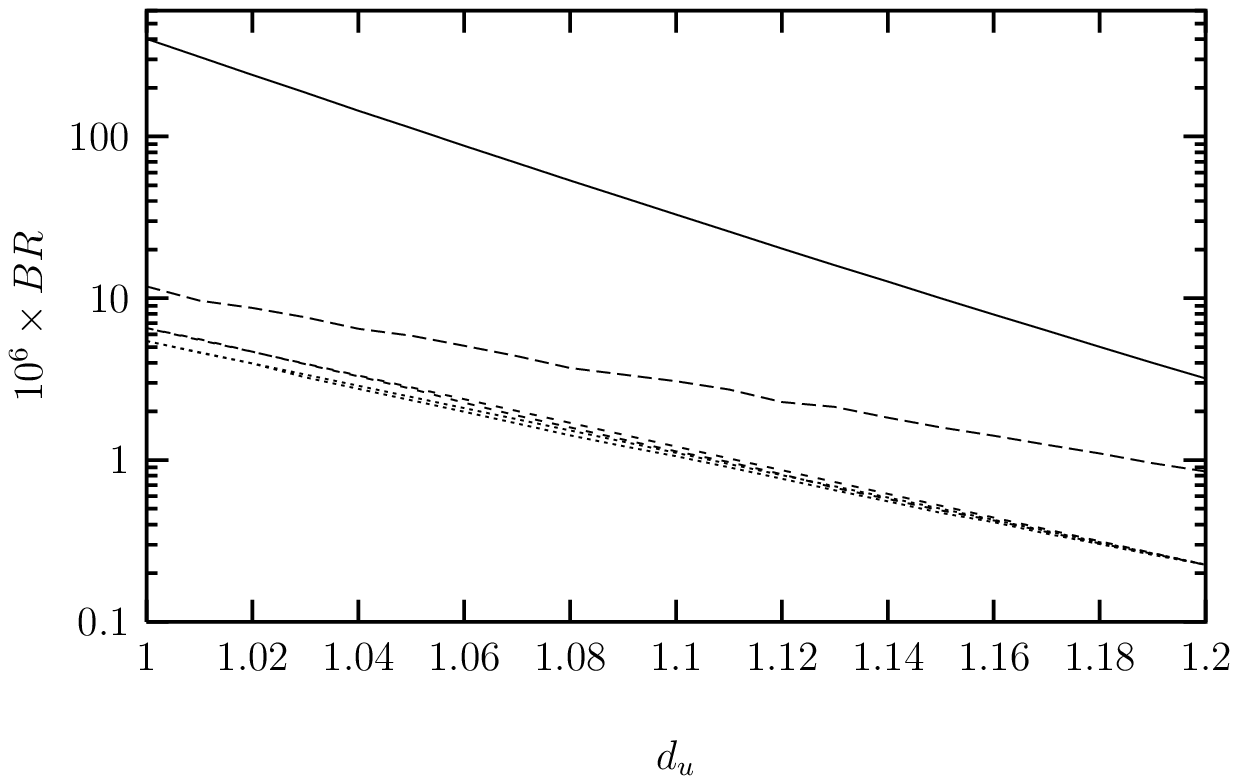} \vskip -3.0truein
\caption[]{The same as Fig.\ref{H0mue120du} but for
$H^0\rightarrow \tau^{\pm}\, e^{\pm}$ decay.} \label{H0taue120du}
\end{figure}

\begin{figure}[htb]
\vskip -3.0truein \centering \epsfxsize=6.8in
\leavevmode\epsffile{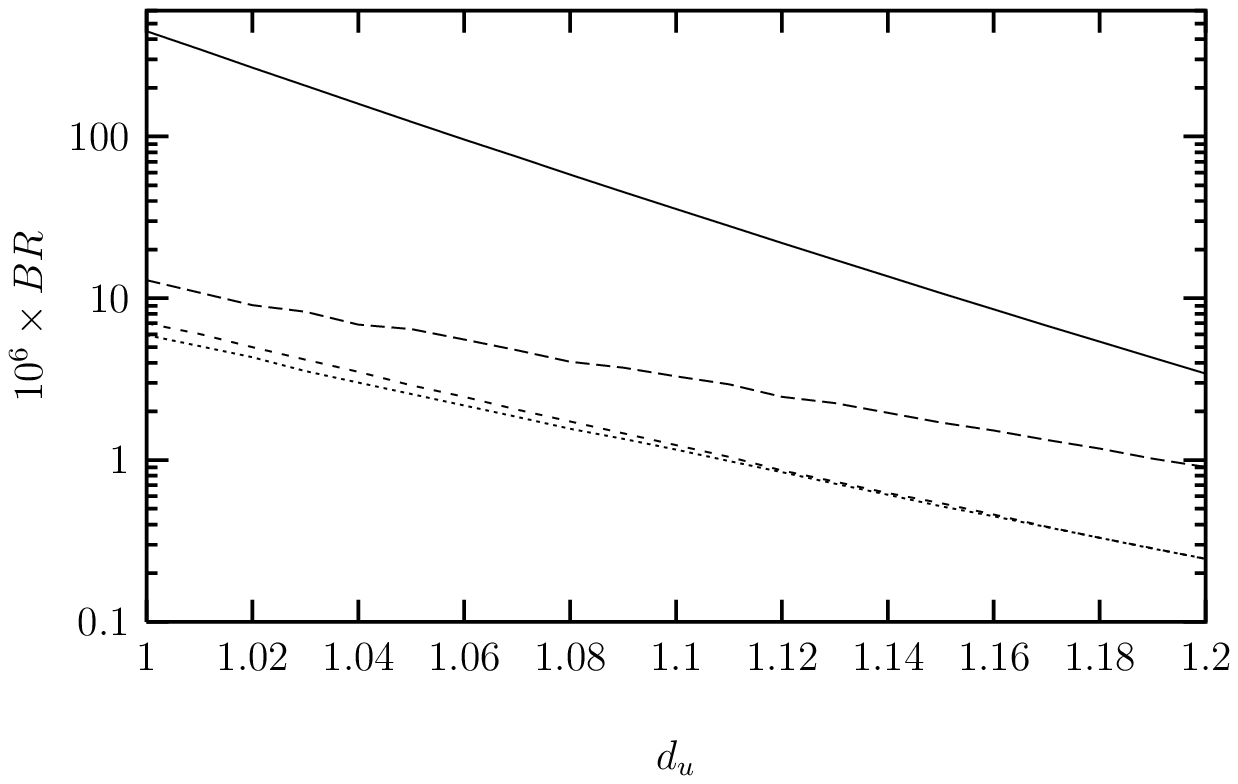} \vskip -3.0truein
\caption[]{The same as Fig.\ref{H0mue120du} but for
$H^0\rightarrow \tau^{\pm}\, \mu^{\pm}$ decay.}
\label{H0taumu120du}
\end{figure}

Fig.\ref{H0taue120du} (\ref{H0taumu120du}) shows the
BR$(H^0\rightarrow \tau^{\pm}\, e^{\pm}\,(\tau^{\pm}\,
\mu^{\pm}))$ with respect to the scale parameter $d_u$, for the FB
couplings $\lambda_{ee}=\lambda_{\mu\mu}=\lambda_{\tau\tau}=1.0$.
Here, the solid-long dashed-short dashed-dotted lines represent
the BR for $\lambda=0.0-0.2-0.5-1.0$.  In the case of $d_u\sim
1.0$, the BR is almost $5.0\times 10^{-6}$ ($6.0\times 10^{-6}$)
for $\lambda \sim 1.0$ and enhances up to $4.0\times 10^{-4}$ for
$\lambda=0$ and $d_u\sim 1.0$. Similar to the previous decay the
mass term leads to a suppression in the BR and the additional term
coming from the $U-U-H^0$ vertex is not enough to enhance the BR
over the numerical values which is obtained for the massless
unparticle case.

\begin{figure}[htb]
\vskip -3.0truein \centering \epsfxsize=6.8in
\leavevmode\epsffile{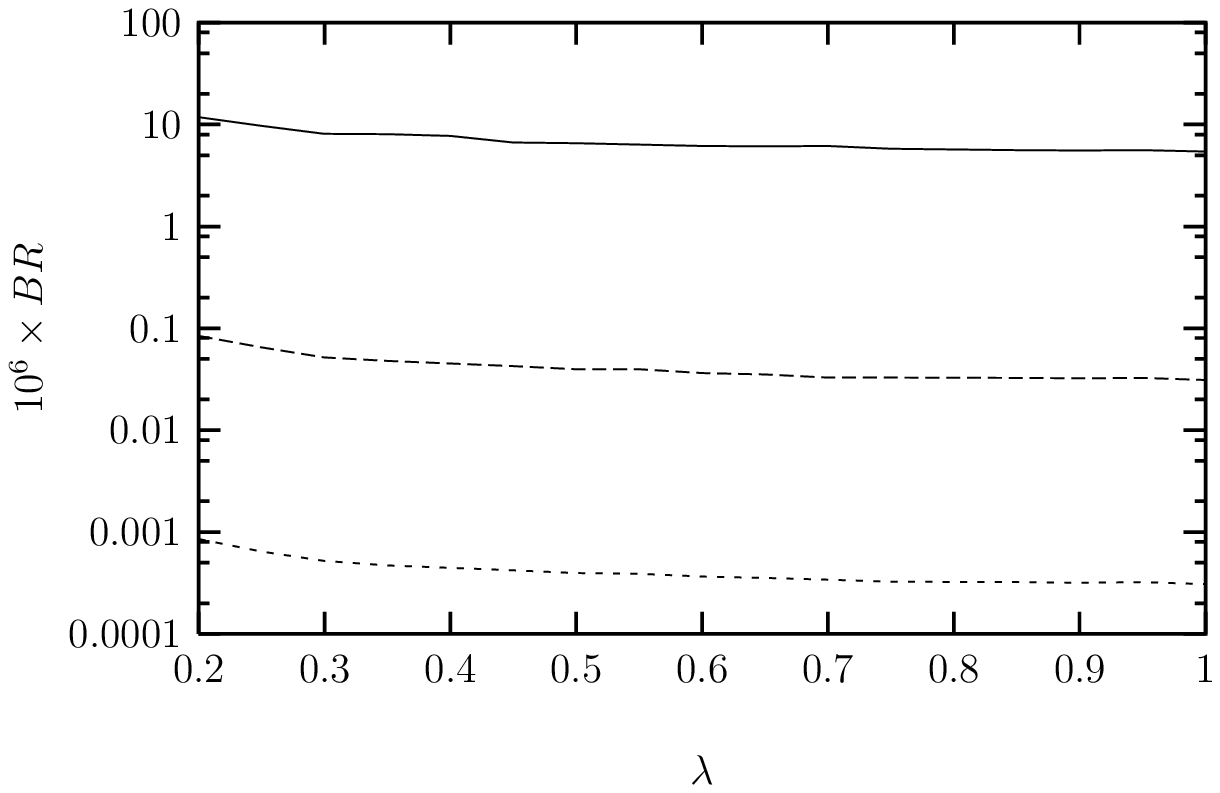} \vskip -3.0truein
\caption[]{The same as Fig.\ref{H0mue120lam} but for
$H^0\rightarrow \tau^{\pm}\, e^{\pm}$ decay.} \label{H0taue120lam}
\end{figure}

\begin{figure}[htb]
\vskip -3.0truein \centering \epsfxsize=6.8in
\leavevmode\epsffile{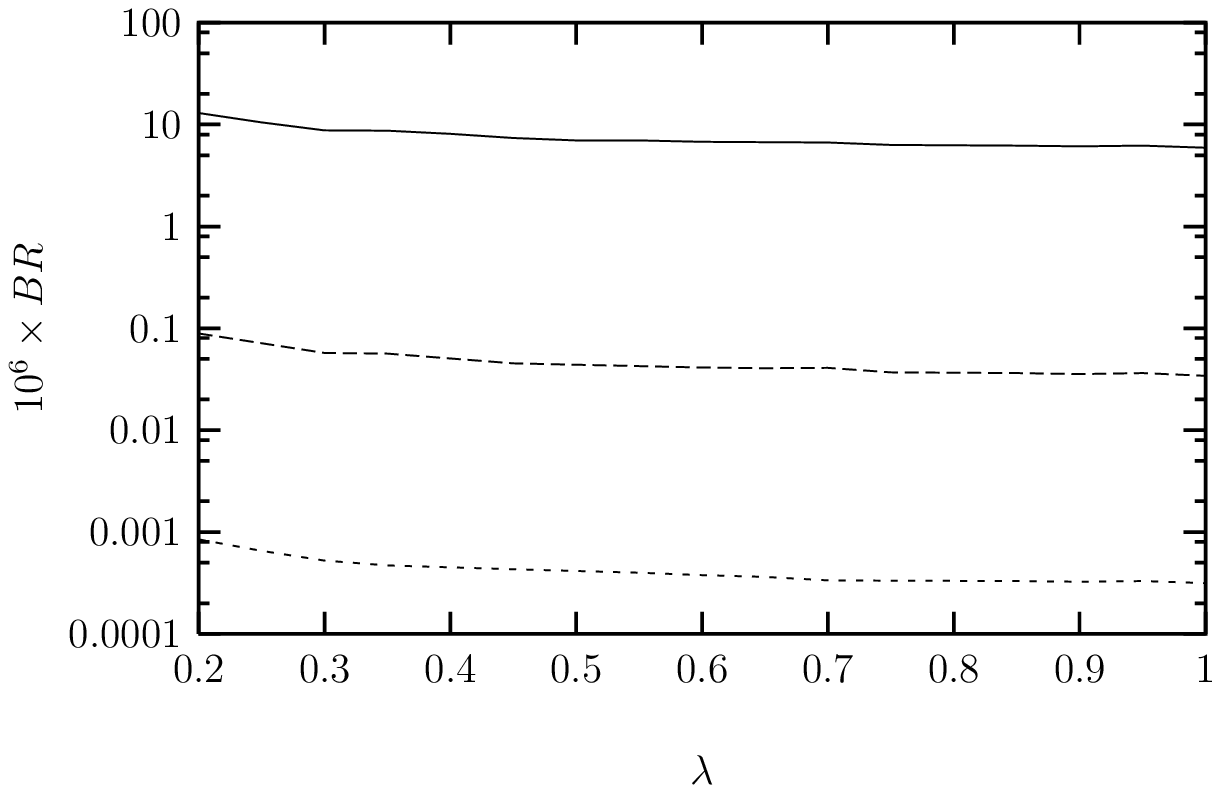} \vskip -3.0truein
\caption[]{The same as Fig.\ref{H0mue120lam} but for
$H^0\rightarrow \tau^{\pm}\, \mu^{\pm}$ decay.}
\label{H0taumu120lam}
\end{figure}

In Fig.\ref{H0taue120lam} (\ref{H0taumu120lam}) we present the
BR$(H^0\rightarrow \tau^{\pm}\, e^{\pm}\,(\tau^{\pm}\,
\mu^{\pm}))$ with respect to $\lambda$ for the scale parameter
$d_u=1$. Here, the solid (long dashed-short dashed) line
represents the BR for
$\lambda_{ee}=\lambda_{\mu\mu}=\lambda_{\tau\tau}=1.0$
($\lambda_{ee}=0.1,\,
\lambda_{\mu\mu}=0.5,\,\lambda_{\tau\tau}=1.0$-$\lambda_{ee}=0.01,\,
\lambda_{\mu\mu}=0.1,\,\lambda_{\tau\tau}=1.0$). It is observed
that the BR is suppressed more than one order in the range $0.0 <
\lambda < 1.0$ and this suppression is strong for $\lambda < 0.3$.
\\ \\
{\Large \textbf{Conclusion}}
\\ \\
As a summary, the mass of unparticle which arises with unparticle
Higgs scalar interaction results in that the BRs of the LFV
$H^0\rightarrow l_1^{\pm}\, l_2^{\pm}$ decays are suppressed. The
BRs are of the order of $10^{-6}$ for $\lambda \sim 1.0$ and
$d_u\sim 1.0$. If the unparticle-Higgs scalar interaction is
switched off unparticle remains massless and the BRs of the decays
studied reach to the values of the order of $10^{-4}$ for FB
$\textit{U}$-lepton-lepton couplings. With the possible production
of the Higgs boson $H^0$ at the LHC the theoretical results of the
BRs of the LFV Higgs decays will be tested and the new physics
which drives the flavor violation, including the unparticle sector
will be searched.
\newpage
\newpage
\newpage
\end{document}